**Reinventing the Arcade: Computer Game Mediated Play Spaces for Physical Interaction**

A. M. Connor and J. Gavin
Auckland University of Technology, Private Bag 92006, Wellesley Street, Auckland 1142, New Zealand.

**Abstract**

This paper suggests that recent developments in video game technology have occurred in parallel to play being moved from public into private spaces, which has had impact on the way people interact with games. The paper also argues and that there is potentially value in the creation of public play spaces to create opportunities to utilise both technology and body for the benefit of community culture and experiences through gaming. Co-located social gaming coupled with tangible interfaces offer alternative possibilities for the local video game scene. This paper includes a descriptive account of Rabble Room Arcade, an experimental social event combining custom-built tangible interface devices and multiplayer video games. The event was designed around games that promoted a return to simplicity through the use of unique tangible controllers to allow casual gamers to connect to the game and to each other, whilst also transforming the event into a spectacle.

**1. Introduction**

The video game industry has grown dramatically over the past decade, cutting into traditional media in participation and revenues as it becomes part of mainstream media culture [1]. This growth can be attributed to a wide range of factors, such as the significant advancement of computing technology and increases in available leisure time, particular with significant decreases in working hours in the 14-24 age group [2]. Whilst this decrease has been accompanied by an increase in time spent in education for this age bracket, overall in the US for this demographic leisure time has increased by 6.2 hours per week for men and 4.9 hours per week for women between 1965 and 2003 [3].

Both game designers and gamers alike believe that gaming is fun [4, 5], though that view is not universally held. Whilst gaming is sometimes (and somewhat naïvely) viewed by the non-gaming public as an isolating activity, it is surprisingly social [6] as the roots of video gaming emerge from a much wider culture of play. However, in video gaming the social element is often related to co-located, parallel gameplay [6] rather than true interactive social play. Social play is often characterised by a degree of frivolity, play for the sake of play where as video gaming is often perceived as play for the sake of competition or power [7].

The nature and value of play is becoming of interest to a range of researchers [8, 9] who have drawn on observations around play theory and the concept that play is a developmental process [10, 11], where arguably frivolous play can be viewed as a more enlightened state of play. It has been observed that play has an elusive value that can be inferred by the observation that many adults pursue activities that have little extrinsic reward with relish and gusto [12]. What is emerging from the literature is that play is not restricted to childhood. Early philosophical notions of play, community and interaction suggest that play is a precursor and principal element of culture [13]. Despite this, it has been noticed that "most American adults have little respect for play, for themselves or, increasingly, for their children" [14]. This presents an interesting state of affairs, where perhaps cultural erosion is occurring over time as people in developed societies focus more and more of their time and energy on work. Csikszentmihalyi [12] suggests that by focusing less on money, power and prestige that it is possible for life to be more meaningful. This is supported by suggestions that public spaces can be used to capture the more spontaneous elements of play and playfulness [15].

Most theories of human play associate play with the freedom of human beings to express themselves openly and to render creatively the conditions of their lives [16], so as culture and society change the nature of play is also changing. The rapid growth of new technologies has made computer video games the normal, to the point where most card or board games also have an electronic version. The Entertainment Software Association (ESA) have reported that 59% of US population play video games [17]. The ESA go on to suggest that 62% of gamers play with other people, either online or in person, and that 47% of gamers play social games. However, ESA also report that 51.9% of all video games sold would be classified as either "Action" or "Shooter" genres.

Whilst the concept of genre in games is considered to be an imprecise and intuitive concept that is impervious to rigorous classification [18], games that are classified in these genres often do not fully explore the diverse possibilities in a broad spectrum of potential gamer experiences. Ravaja et al. [19] have investigated the spatial and emotional responses of gamers playing against computer opponents, strangers and friends and concluded that playing against a friend elicited greater spatial presence, and self-reported and physiological arousal compared to playing against a stranger. In addition, Yee [20] observes that online gamers spend on average 20 hours a week in online games, and many of them describe their game play as obligation, tedium and more like a second job than entertainment. When combined with the observations of Ravaja et al. [19] this suggests that there is an opportunity to reconsider the nature of video games to bring back the fun, and that the design of new games and interfaces in a social context can break the behaviour conditioning principles embedded in many video games that inherently train players to become better "game workers".

This paper outlines the design of a social play event that is based around challenging common perceptions related to video games and taking the concepts of gaming from co-located console play to one of more overtly physical, cooperative social play. This is not to suggest that there is anything inherently wrong with existing modes of video game play. The project was undertaken to explore different modes of game play that might appeal to a wider audience.

The project actively rejected "polish" in favour of simply "being done", particularly in terms of the interfaces developed though also the games themselves. The focus was to produce an outcome that was sufficiently functional to be enjoyable, retaining any quirks obtained during the creation and not utilising techniques or materials normally associated with mass produced commercial products. This was driven by a philosophy to re-use, recycle, and re-purpose instead of buying new, as part of a self-awareness of an over-consuming society.

## 2. Background & Related Work

Ever since they were popularised by the emergence of Pong in 1972, video and computer games have become increasingly part of mainstream entertainment culture. In today's society, video games are not just played on computers and game consoles but also handheld devices and cell phones. Because of the ubiquitous nature of these devices, games are no longer just played at home and at arcades, but are also played at work, at school, on public transport, and virtually anywhere that an electronic device can be operated.

The amount of time spent playing games has increased over time [21] and it is common for children and adolescents to play more than 20 hours each week with 40 hours of gaming not being uncommon among young males [22]. In some cases it has been observed that gaming can become an addiction [23, 24] with it being observed that so called "pathological gamers" spent twice as much time playing as non-pathological gamers and often received poorer grades in school as well as exhibiting attention problems [25]. The vast growth in gaming has driven considerable research which has examined potential positive and negative effects of playing various types of video games. Much of this work has focused on the detrimental effects of playing violent games [26] or further exploring a negative association between time spent playing games and school performance [27]. However, gaming does present a particular dilemma as there is much research that also emphasises the positive value associated with educational games [28], that games do have the potential to increase prosocial behaviour [29, 30] and that exercise games are an attractive form of physical activity [31, 32] and are arguably able to be fun and engaging for a wide demographic of player [33]. It seems that the impact of gaming depends very much on the game itself, the nature of play and the play environment. In the past forty or so years in which video gaming has become popularised, all of these elements have seen considerable change.

Despite the growing ubiquity of gaming and its significant influence upon youth culture, relatively little is known about the industry's evolutionary path and internal dynamics [34]. Ayoma & Izushi identify that the video-game industry comprises large console manufacturers, video-game publishers, and different sized video-game development organisations [34]. This structure is representative of the whole digital interactive entertainment industry today, which is controlled by a relatively small number of global corporate developers, publishers, and distributors [35]. In terms of the video-gaming industry, the most lucrative aspect of the games industry is an outcome of the competition between different gaming consoles [36]. Some of the successes of the industry are derived from the forging of alliances between console manufactures and major software publishers [34]. In contrast, there is a growing independent game developer community [37] that exists outside of the major players, and the interplay between the various elements of the industry produce a system of production [38] that has grown increasingly more complex over time.

The history of video gaming has been described in detail by many authors [39] so will not be considered in detail in this paper. However, reflection on the turbulent history of the gaming industry provides insights in to the nature of play. Williams argues that the early 1980s were a crucial turning point in the social history of video game play that saw an erosion of what began as an open and free space for cultural and social mixing [40]. The history of video gaming can be summarised as slow adoption during the 1970s leading to a massive spike in popularity during the Atari heyday of the early 1980s, followed by the collapse of that company and the industry's eventual revival in the late 1980s by Nintendo. Williams argues that "video games helped usher in a new kind of consumer, one increasingly aware of new tools and new possibilities. Consumers were beginning to embrace home computers, compact discs, and the concept of digital systems as convenient and powerful entertainment tools" and this influenced where play was conducted [40].

The arcade establishment was the primary medium for the video game experience during the 1970's and 1980's, the "golden age of arcade video games" [41]. Despite the attention mandated by the video game screen, early arcade games embraced multiple dimensions of physical experience, such as the 1975 eight-player game Indie 800, which had a steering wheel and two pedals for each player [41] as well as the screen. These video arcade machines were an offshoot from earlier mechanical games, such as pinball. Designers were therefore attentive to the tangible interaction aspects of the game. Such machines, and their earlier mechanical counterparts, exhibited a sense of spectacle in the arcade and some authors have observed that the early arcades were a very social phenomenon [42]. The desire to recreate this sense of spectacle around a social phenomenon is one of the motivating factors for the work outlined in this paper.

The late 1980s also saw the beginning of play moving from public to private spaces. Throughout the 1980s, a combination of economic and technological forces moved play away from social, communal and relatively anarchic early arcade spaces, and into the controlled environments of the sanitized mall arcade (or "family fun centre") and into the home [40]. This was in part driven by the uptake of the home computer and game console in the 1990's, which reduced much of the social and mechanical aspects of computer gaming through a process of downsizing arcade based machines into smaller units played at home. One of the side effects of this transition is that the large multiplayer elements of traditional arcade games were often replaced by at least some emphasis on single-player gameplay. On this point, Salen and Zimmerman [43] describe single-player gaming as an anomaly in the rich history of games.

While many modern video games embrace multiplayer modes through computer networking, screen-based gaming with a standardised button interface continues today as the main adult experience of games [44]. It has been argued that the input button that is so central to video gaming is impeding the development of the medium [44] because the button "[disregards] the bodies abilities" and permits the player to "forget about the physical device". Researchers have issued a call to arms to abandon the button as soon as possible and replace it with more natural interfaces [45]. This paper argues that the button itself is not impeding the medium's development, but conventions of usage surrounding the button do impact development. Designers and players have utilised technology to provide the cheapest and most efficient route to gratification, based around the use of fine motor skills and small motions to facilitate rapid action. The social and physical implications include reduced exertion and less face-to-face bonding with others when compared to the large numbers of traditional games in which the abilities and idiosyncrasies of the body are essential to the play and enjoyment of the game [46]. Studies investigating tangible interfaces for video games involving the use of gross motor skills show the opposite [47].

Whilst play is inherently social in its nature, research has shown that playing online multiplayer games does not produce the same prevalence and extent of social activities as might have been expected [48]. However, Mueller, Gibbs & Vetere [49] argue that so called "exertion games" are emerging that have not only have potential health benefits because they promote exercise but they can also facilitate social play between players and that social play can improve participation in such games that may not be as appealing to an individual alone.

Whilst Mueller et al. [49] make these observations about exertion games, their work mostly focuses on gameplay that is not co-located but instead facilitated by computer networks. In contrast, this paper argues that there is an increased benefit in developing co-located games that have elements of exertion play and unique interfaces to encourage greater interaction. There is therefore a benefit in terms of more social and overtly physical play, and this paper proposes a return of the somewhat anarchical arcades of the 70s and 80s that differed from the later sanitized mall [40]. The stereotype of gamers as antisocial creatures needs to be challenged and the emergence of LAN parties [50] suggests that gamers are in a way more tribal than solitary and thrive on the social aspects of play. This is borne out by the emergence of e-sports, in which competitive gameplay borrows forms from traditional sport [51]. Some authors go as far as to suggest that gaming is often as much about social interaction, as it is about interaction with the game content [52]. In this context of socially situated play, this paper argues that whilst there is value in playing traditional video games that there is also a place for community based play events that are designed to increase interactivity through a greater physicality of play. Such events have the potential to promote the development of the gaming medium by reaching out to a broader spectrum of participants.

This view is borne out by the emergence of other localised movements countering the potential isolation and sterilisation created by the use of modern technologies. For example, the New York collective Babycastles (created by Kunal Gupta and Syed Salahuddin in 2010) provides a local play space to showcase artistic, independently-created video games and interfaces, alongside visual artists, installation artists and musicians. Similar projects are springing up around the world, including the LA Game Space, an inclusive workshop and gallery for people to explore unconventional possibilities related to games. In New Zealand, Guerilla Playspaces is an Auckland-based project that encourages public play through artefacts and installations. As Pasternack affirms, "the patrons of... these independent communities, are, in one way or another, striving to experience something new; something that can't be bought in a store, but that is available for anyone to see and hear if they look in the right places. Just like indie music, the independent gaming scene is trading in neat, mass-produced convenience for a rough-hewn, playful provocation" [53]. This paper describes a project that reintroduces both social connection and increases physicality in the gaming experience and purposefully provides such a rough-hewn and playful provocation.

**3. Project Overview**

Rabble Room Arcade [54] was a project conducted in 2013 by students at Auckland University of Technology. The project Rabble Room Arcade was the identity created for the social play event held in October 2013, and the choice of the word 'rabble', meaning the common people; disorderly crowd; or a "boisterous throng of people" intentionally focussed the context on community and agitation.

The team set out to showcase independently-developed games, embracing the absurd and overtly physical, for the purpose of exciting a local cultural experience. The intention is to promote a more ludic aspect of gaming that promotes a return to simplicity through using unique tangible controllers to allow casual gamers to connect to the game and to each other. By reconsidering the value of play in public and urban spaces [15] it is possible to perceive a number of ways that such an event can be conceptualised. McGonigal [5] advocates gaming for change, especially in the face of global problems, arguing that "games are a sustainable way of life". Playing games with others eases our suffering, conserves resources, and facilitates participation in supportive and coordinated communities. Active participation "increases the likelihood that one will learn from the video game due to greater identification and immersion" [55]. One form of active participation is when the body is engaged in play. Tangible interfaces allow "physically engaging experiences with technology" [56] and novel tangible interfaces have an intuitive appeal [57]. One of the goals of the project was to allow such engaging experiences to emerge for a wide range of participants.

The project also included a social purpose to positively contribute to the wider gaming community, by connecting people through the experience of play and promoting a sense of community amongst the participants in the event. This was achieved by seeking out the unusual; making social, political, and environmental statements; and imparting knowledge to the community throughout the development of the project. In this context, unusual is defined as being more than games with quirks and differences. One of the goals of the project was to not just upset video game tropes but also quietly poke fun at a small part of the status quo of Western culture through an expedition into the unconventional. This expedition explored tangible interfaces and pushed boundaries of the notions of conventional. The overall outcome of the project was to convey a message of exertion for play instead of exertion for profit, and issue an appeal to laughter and fun with a determination to change the city and community for the better.

### 3.1. Game & Interface Development

The development of the arcade event consisted of three intertwined strands, namely the organisation of the space, the elicitation of games from local independent developers and the development of interfaces for each game. Rabble Room Arcade featured eight very different games as highlighted in Table 1.

Table 1. Game Descriptions

| Game | Description |
| --- | --- |
| *Double Shovel* by Jeff Nusz | A game where two players would cooperatively shovel grain into a chute to trigger events like feeding a child or cleaning up a kitchen. |
| *Elevator* by CyrilQ Studios | A two player competitive game with cranks as input devices that have to be operated as fast as possible to make the game character go up an elevator as fast as possible while avoiding virtual objects being thrown at them. |
| *Space Octopus Mono* by Matthew Gatland | An 8-bit style arcade game where the players control the horizontal position of the spaceship via wooden sliders on wooden rails. |
| *Off Da Railz* by Vox Populi: | A game where the player controls a train with a wooden board that has tilt sensors for direction and speed control. |
| *CatManDudu* by Emile Drescher and Tom Tyer-Drake | An experimental game controlled by two foot-operated buttons for direction and a toilet chain switch for triggering "shots". |
| *Word Wars* by Jenna Gavin and Tom Tyer-Drake | A competitive game in which up to eight players form words by "grabbing" letters that appear on the screen by pushing a single button. |
| *Fruit Racers* by Jenna Gavin and Tom Tyer-Drake | A four player competitive game with rotary encoders as input devices to control the direction of fruit on the screen in a race setting. |
| *Shadow Showdown* by Matthew Martin, Jenna Gavin, and Daniel Cermak-Sassenrath | A cooperative game where one or more players have to match silhouettes on the screen by creating silhouettes with their own body/bodies. |

Of the eight games showcased, only two games were fully developed by the event team. The interfaces for the remaining six were developed by the event team, however in most cases the games themselves were developed either wholly by or in conjunction with external contributors. For the games described later in this paper, the marrying of interfaces with games was an iterative process where game concepts were developed and the interfaces prototyped and tested, leading to changes in the game itself before finalising the interface. However, many of the other interfaces were developed for emerging games that utilised either keyboard, mouse or joystick based controls. To simplify the interaction, the Lightweight USB Framework for AVRs (LUFA) framework was used to emulate these devices on an Arduino microcontroller. This removed the necessity for the game developers to accommodate a specific interface in their game implementation, allowing them to instead focus on the gameplay. The game developers were briefed to produce games that could be competitive or collaborative.

In terms of the development of the interfaces, one of the main techniques employed was to explore interfaces that opposed optimised efficiency. This was done not only for the purpose of disrupting expectation (and thus encouraging active, divergent thought), but also to even out the "playing field", so that the games did not privilege those that had trained for hours on standard interface devices. This "gestural excessiveness, as a showy form of inefficient gameplay, represents a refutation of hardcore instrumental play" [58]. The technique of designing for inefficiency works very well for social spectacle but may degrade with long-term play. This paper proposes that inefficiency of the interface and interactions should be considered in a light-hearted and social environment; for "especially in regards to party and street games, public spectacle comprises the heart and soul of what those activities are" [58].

Figures 1-3 highlight a number of the interfaces developed for the games, with Figure 1 showing the interface for Double Shovel, Figure 2 showing the rotary encoders used in the game Fruit Racers, and Figure 3 showing the ultrasonic sensors and sliding rails used in Space Octopus Mono. This is a representative sample of the diverse range of interfaces used.

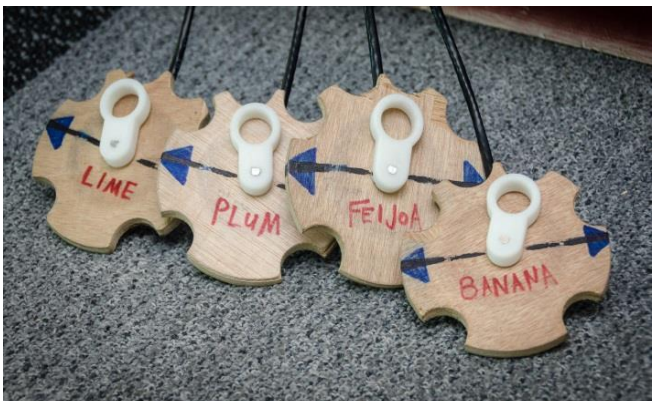

**Figure 1.** Interface for Fruit Racers

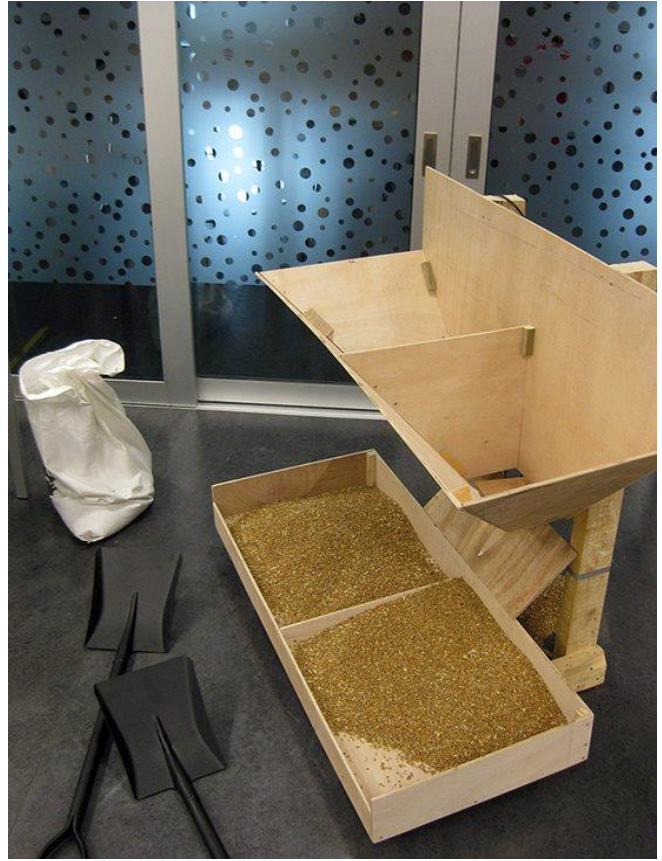

**Figure 2.** Interface for Double Shovel

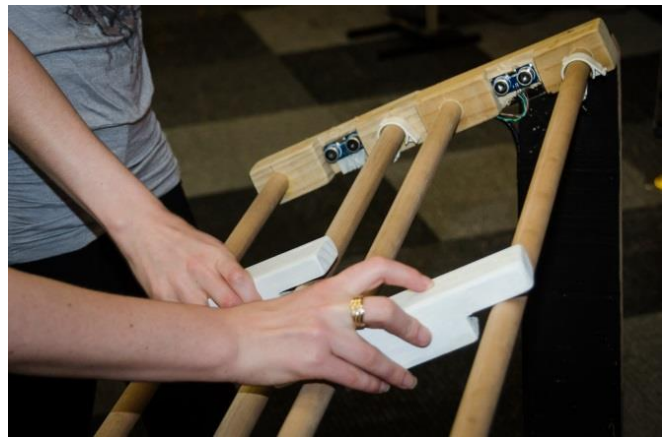

**Figure 3.** Interface for Space Octopus Mono

A full discussion of the games, the interfaces and the mediated play environment they produce would be overly detailed for the purpose of this article. A brief discussion of a number of the games will provide sufficient insight to the design intent of the developers in terms of how the games are intended to promote both gross physical movement and social interaction. These games are selected because the interfaces share a common design goal which was the creation of proximity between players. The decision to promote such proximity is not entirely grounded in a theoretical framework, though it has been observed that physical proximity allows for a more intense and multi-sensory awareness during gameplay [51].

**Shadow Showdown**
The development of Shadow Showdown [59] has been informed by the transformation of the game Twister in to three dimensional space. As a game, it is designed around whole body interaction and is intended to challenge boundaries of space and the body. The game focuses on interaction in real world space within two contexts facilitated by a Kinect. The two contexts are firstly the relationship between players and the screen, and secondly the rapport amongst players. Gameplay is casual and intended for a general audience. Players use their bodies to mimic and fill in shapes displayed on a screen. The silhouettes of players are also displayed on screen, converging with the shape. After 15 seconds, a snapshot is taken and correct silhouette coverage is awarded percentage points. Players can either participate in teams or alone, cooperatively or competitively. The different shapes embedded in the game are shown in Figure 4.

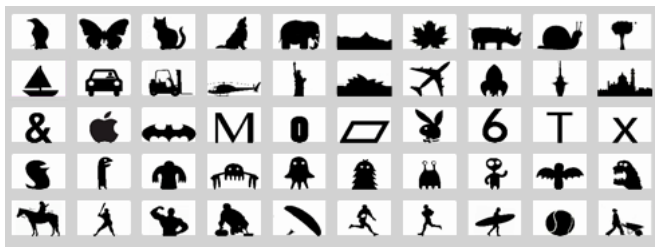

**Figure 4.** Game Shapes in Shadow Showdown

Figure 5 shows the game undergoing initial testing, and is composed of three parts. The first is the image that the players are attempting to fill, the second is the players themselves and the third is the difference map used to compute the score.

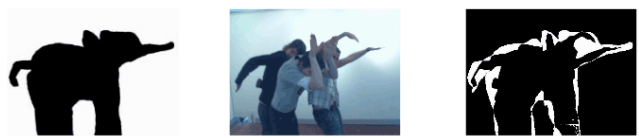

**Figure 4.** Game Shapes in Shadow Showdown

In terms of Rabble Room Arcade, the game stands out as the only game included where the human body is the controller. The design of the game is such that when team play is undertaken that proximity between the team is unavoidable.

Since the introduction of the Nintendo Wii, the proliferation of ubiquitous gaming platforms, and the introduction of social games, there has been a steady change in who interacts and how with games. Games such as Shadow Showdown are often referred to as "casual games" and are now played by players of all ages who have no need to possess an intimate knowledge of video game history or devote weeks or months to play [60]. Juul describes this as a reinvention of video games and suggests that "many of today's casual game players once enjoyed Pac-Man, Tetris, and other early games, only to drop out when video games became more time-consuming and complex" [60].

Whilst Shadow Showdown embraces this return to simplicity through using the body as a controller, other games showcased at the Rabble Room Arcade utilised unique tangible controllers to allow casual gamers to connect to the game and to each other through a process of exploration and communication. In this game, players can become more connected to each other through the need to collaborate in order to produce a good facsimile of the shadow.

**Word Wars**
Of particular interest is the game "Word Wars", conceived in 2013. The game springs from a minimalist design perspective, namely, the investigation of game mechanics that arise if each player only has one button. It has a simple rule set, is casual yet tense, and encourages a tight social experience around a waist-height cabinet. It is built in Processing for the purpose of receiving multiple button presses through Arduino, where a standard keyboard will limit simultaneous key presses to six. Gameplay in Word Wars is based around completing an English word more than three letters long. Each player pushes their button to try and grab the letter in the middle of the shared screen. The design of the mechanical elements of the game are purposefully such that the game occupies a small footprint, requiring players to be in close proximity and conjoined in a competitive but fun atmosphere naturally focused on the game. The interface therefore purposefully creates a closed space, where the proximity of the players creates the perception of a "huddle" of intimate privacy when viewed from outside the game. The game is shown in Figure 6.

In its own way, this game challenges the pervading view that the "button" is impeding game media development and rejects the call to "discard the button in favour of natural interfaces". It clearly demonstrates that the simple button can in fact facilitate more social game play and be used in innovative and exciting game designs. This suggests that perhaps the humble button is not a major issue, but instead the lack of creativity in designing play in a fun and engaging manner that may include new ways of appreciating the button. This is supported by the attestation of Juul [60] who suggests that "it is only by understanding what a game requires of players, what players bring to a game, how the game industry works, and how video games have developed historically that we can understand what makes video games fun and why we choose to play (or not to play) them.".

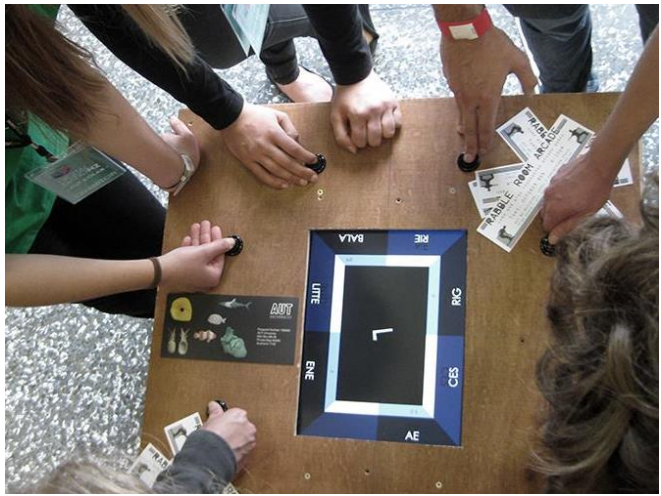

**Figure 6.** Word Wars

**Fruit Racers**
The Fruit Racers game provides another example of how the games in the arcade are based around the principles of reinventing video games using alternative interfaces. In this case, the controller (as shown in Figure 2) consists of two tilt sensors and a rotary encoder placed upon a board that is easy to hold. Just as with using the body as a controller, the installation of new tangible interfaces creates a more general appeal and a level playing field where experienced gamers and casual gamers can participate in a game with no disparity arising from familiarity with the controller, thus promoting a greater degree of closeness between the players.

Gameplay is very simple, owing to the nature of the controls. Various fruit race to touch the checkpoints in a derby arena. Tilting the sensor to the left or right causes the fruit to turn, whereas the rotary encoder is used to accelerate and decelerate in a given heading.

Although still in a very rough form, the game is intended to show that novel controllers can create a casual and fun atmosphere amongst a range of different types of gamer. As with the other games highlighted in this paper, physical proximity is encouraged through the purposeful selection of different lengths of the cable attached to the individual controllers.

**4. The Rabble Room Arcade**
Whilst the philosophical underpinnings of the Rabble Room Arcade have already been discussed, in practice the Arcade was designed as a pop-up social play space that brings the community together through an atmosphere of relaxed, unique fun. The development of the event was a final year project for students who were looking for an opportunity to explore their creative craft while organising an event centred on public participation. In this regard, the project was undertaken without a strong theoretical basis, instead focusing on an inductive approach where observations from the event. The event was visited by more than 100 people and also featured on a current affairs TV show, Seven Sharp.

The intention for the arcade was to feature locally-developed, multiplayer video games that blend the tangible with the digital, allowing the public to experience innovation, imagination and play in a social setting. The ethos of the development of the arcade was grounded in a view about how society and culture are impacted by technological change. As society has moved to embrace technology, there seems to be a counter-movement away from potential isolation. Play and expression counter apathy and depression; impression without expression leads to depression. The nature of expression is acutely described by Hahn as "We are meant to express the way we feel about life. It's like breathing:

inhale the experiences of life, exhale how you feel about them" [61]. Calleja [62] goes on to explain how games provide expressions of agency to the player through decisions, meaningful action, and by seeing the results of his/her intentions. Rabble Room Arcade is therefore an expression through the act of creating and designing games and art. Figure 7 shows gamers at the arcade interacting with a number of the games.

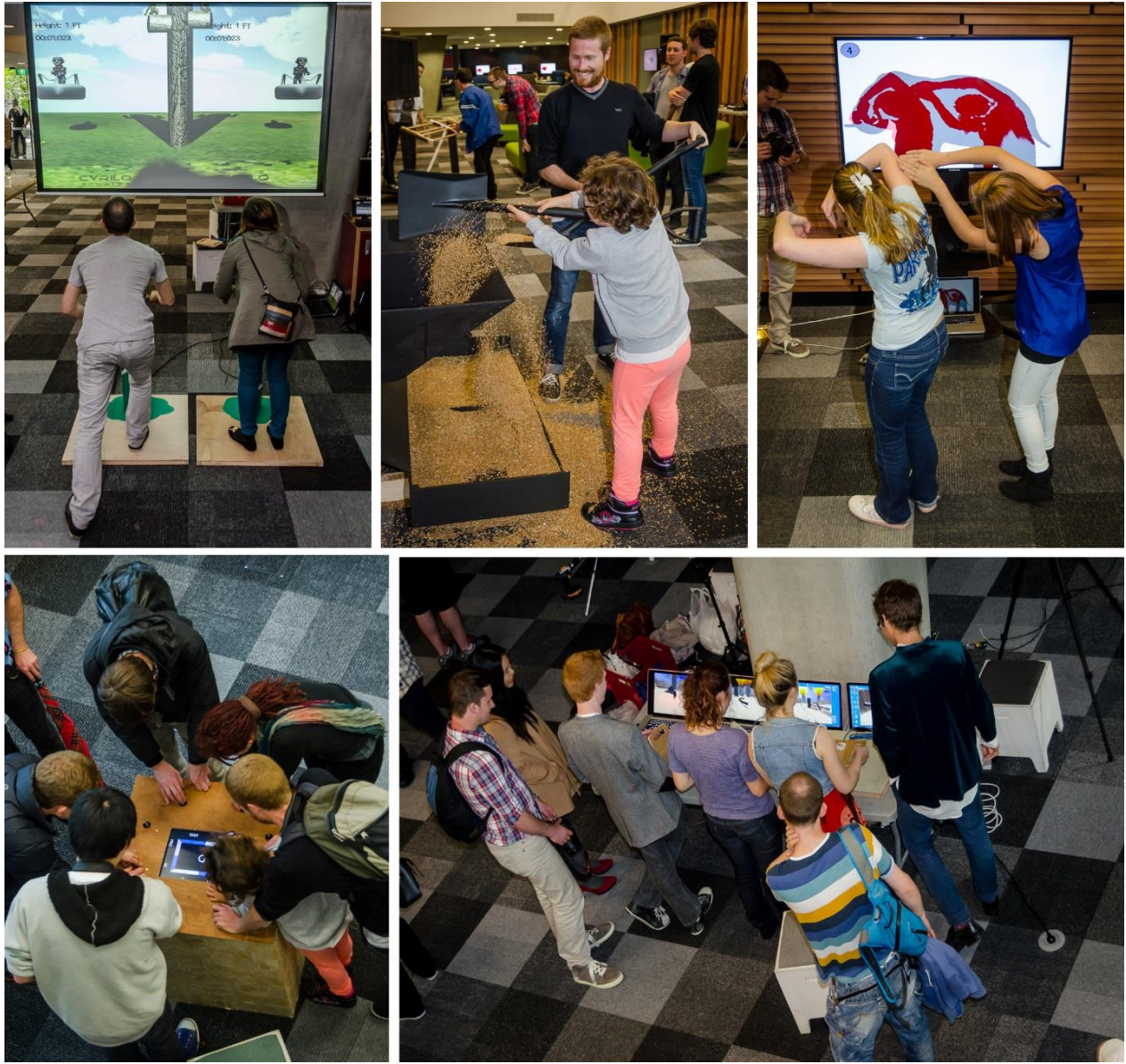

**Figure 7.** Gameplay at the Rabble Room Arcade

The Rabble Room Arcade project has proven itself to be a local flavour in an exciting global movement towards more community-focused gaming, one in which alternative tangible interfaces and experimental ideas flourish. Games do not exist in a social vacuum and whilst some may argue that games themselves can become a "third place" [63], a place that is neither home nor work, others argue that the creation of new "third places" has the potential to improve the quality of life in a community [64] that is in line with the current direction of the local council's development plan.

It has been noted that games culture is at a particular defining moment for social reasons as well as technological ones, as well as a general trend that has seen the transition of public life from common spaces to private ones. This trend is exemplified in the move of game play from arcades to homes [65]. However, the Rabble Room Arcade experience suggests that there is potential for a bright future for the arcade as people rediscover the joy of playing games for the sake of social interaction in public spaces. It brings together game developers with players, particularly in a way that introduces non-gamers to the potential that gaming can be a socially cohesive force.

The event has provided insights into what technical, organisational, and social hurdles are faced in terms of developing social play spaces. Reflecting on the process of organising the arcade and the outcomes provides emphasis on the importance of engagement. In this context, engagement can refer to engagement with "the idea of the event" as well as engagement with the event itself. Engagement with the idea of the event cuts across multiple facets, and the first interesting point to consider is the number of games contributed by game studios or individuals outside of the event team.

Given the relatively short timescale for development, this suggests that the gaming community is inherently social and is looking for opportunities to engage in unique, sociable playspaces. Game developers were excited by the prospect of showcasing their games and embraced the playfulness of the event. It is encouraging that the spirit of the anarchic arcades of the 1970s is still present in the game development community. Evidence of engagement with the event was also positive, with over 100 attendees all of whom embraced the alternative interfaces and clearly identified with the makeshift and local spirit of the event.

**5. Conclusion**
This project offers a viable alternative for video game culture by altering the traditional button interface and content of gaming. In essence, the project is not a research project per se (though it is informed by research) but one of activism to promote change. The project explored interfaces that oppose optimised efficiency and ease of use, and advocated for games that were independently-developed and out of the ordinary. The project embraced gaming, the absurd, and minimalism in the pursuit of reinventing the spectacle of the arcade.

In some regards the project is an instance of rebellion against the status quo as an affirmation of freedom, self-expression, and equality. Authority structures and commercial interests dominate and dictate society, including the big-dollar industry of video games. In a very general sense, change of values in relation to culture, sustainability, and any number of topics can come from actively participating in a process, idea or activity. The Rabble Room arcade was intended to engage with the community, however the level of active participation from the developers and attendees spawned interest in the event from the local media, which provided an opportunity to raise questions regarding the nature of computer games and their role in society to a much wider audience.

Tangible interfaces have had a long history in video gaming, especially in the mechanical cabinets and the arcade machines pre-1990. Relatively recent advances in technology, such as 3D printing, have enabled the layperson to create functional prototypes with ease, using these developments to explore unique, independent, and overtly physical video games. Community based play events can embrace increased physicality of play as a means of increasing social engagement and promoting the development of the gaming medium. However, the possible advantages of gaming as promoted in this work are not restricted to social engagement. There is potential to effect change in wider applications such as health and education, particularly across the full spectrum of age.

The techniques used in the conduct of this work were focused on fast development of working prototypes in a practice-based method to see what would emerge. This approach has its limitations, particularly in terms of ignoring an established body of knowledge in terms of embodiment, materiality and tangibility that could be used to develop more informed outcomes. Such theoretical considerations will be embraced in future work.

Inefficiency of interface and interactions help create a sense of spectacle, and are well-received in light-hearted social settings. The Rabble Room Arcade event demonstrated local acceptance and engagement with unusual tangible interfaces which, as anecdotal evidence suggests, can increase the levels of fun and engagement across a wide range of participants. This has been evidenced by the enthusiasm of the local independent game development community to be involved in the event, as well as the large number of attendees from all walks of life.